\documentclass[conference]{IEEEtran}
\IEEEoverridecommandlockouts
\usepackage{cite}
\usepackage{amsmath,amssymb,amsfonts}
\usepackage{algorithmic}
\usepackage{graphicx}
\usepackage{subcaption}
\usepackage{textcomp}
\usepackage{booktabs}
\usepackage[table,xcdraw]{xcolor}
\usepackage[normalem]{ulem}
\useunder{\uline}{\ul}{}
\usepackage{caption}
\usepackage{hyperref}
\usepackage{pifont}
\usepackage{url}
\usepackage{multirow}

\hypersetup{hypertex=true,
colorlinks=true,
linkcolor=[RGB]{105,33,106},
anchorcolor=[RGB]{105,33,106},
citecolor=[RGB]{105,33,106},
urlcolor=[RGB]{0,76,129}}
\usepackage{svg}
\def\BibTeX{{\rm B\kern-.05em{\sc i\kern-.025em b}\kern-.08em
    T\kern-.1667em\lower.7ex\hbox{E}\kern-.125emX}}

\begin{document}

\newcommand{\method}{RePart}
\newcommand{\submethod}{RePart-mini}
\newcommand{\maxhop}{\textsf{Max Hop}}

\title{\huge \textbf{\method: Efficient Hypergraph Partitioning with Logic Replication Optimization for Multi-FPGA System}\\
}



\author{%
  \IEEEauthorblockN{%
    Zizhuo Fu\textsuperscript{1,3,5\dag},\;
    Yifan Zhou\textsuperscript{1,3,5\dag},\;
    Zhaoxin Lu\textsuperscript{1,5\dag},\;
    Guangyu Sun\textsuperscript{1,2,4},\\
    Runsheng Wang\textsuperscript{1,2,4},\;
    Meng Li\textsuperscript{3,1,2,4*},\;
    Yibo Lin\textsuperscript{1,2,4*}%
  }\\[-2ex]
  \IEEEauthorblockA{%
    \textsuperscript{1}School of Integrated Circuits, \textit{Peking University}\;\;
    \textsuperscript{2}Institute of Electronic Design Automation, \textit{Peking University}\\
    \textsuperscript{3}Institute for Artificial Intelligence, \textit{Peking University}\;\;
    \textsuperscript{4}Beijing Advanced Innovation Center for Integrated Circuits\\
    \textsuperscript{5}School of Electronics Engineering and Computer Science, \textit{Peking University}
    \thanks{\dag\ These authors contributed equally to this work. * Corresponding authors: Yibo Lin (yibolin@pku.edu.cn) and Meng Li (meng.li@pku.edu.cn).}
  }%
}

\maketitle

\captionsetup[table]{
  labelsep = colon,
  justification = centering,
  singlelinecheck = off,
  labelfont = bf,
  textfont = normalfont,
  name = Table,
  skip = 5pt
}

\begin{abstract}
Multi-FPGA systems (MFS) are widely adopted for VLSI emulation and rapid prototyping.
In an MFS, FPGAs connect only to a limited number of neighbors through bandwidth-constrained links,
so inter-FPGA communication cost depends on network topology.
This setting exposes two fundamental limitations of existing MFS-aware partitioning methods:
conventional hypergraph partitioners focus solely on cut size and ignore topological structure, and they leave substantial FPGA resources unused due to conservative balance margins.
We present \method, a fully customized multilevel hypergraph partitioning framework for MFS
that integrates logic replication with topology-aware optimization.
\method~introduces three coordinated innovations across the multilevel pipeline:
FPGA-aware dynamic coarsening, heat-value guided assignment, and replication-deletion supported refinement.
Extensive experiments on the Titan23 and EDA Elite Challenge Contest benchmarks show that
\method~reduces total hop distance by 52.3\% on average over state-of-the-art hypergraph partitioners
with a 11.1$\times$ speedup, and outperforms the EDA Elite Challenge winners. Code is available at: \url{https://github.com/Welement-zyf/RePart}.
\end{abstract}

\section{Introduction}

Field-Programmable Gate Arrays (FPGAs) are widely used in logic verification, where FPGA prototyping enables high-speed, cycle-accurate validation of complex designs~\cite{Ray2003HighlevelMA}. 
Recently, FPGAs have also attracted growing interest for deploying large language models, which often need to be distributed across multiple devices due to their high compute and memory demands~\cite{dfx}.
As design complexity increases, the demand for Multi-FPGA Systems (MFS) has grown substantially~\cite{TESSIER2008637}. 
In such systems, how design components are partitioned and mapped to individual FPGAs largely determines overall performance.

Design circuit netlists can be represented as hypergraphs, naturally formulating the problem as hypergraph partitioning~\cite{646609}. 
Hypergraph partitioning is NP-hard, and has been extensively studied. 
Early approaches such as the KL~\cite{6771089KL} and FM~\cite{1585498FM} algorithms introduced iterative node-swapping techniques. 
Later advancements including hMETIS~\cite{748202hMetis}, PaToH~\cite{C2011patoh}, and KaHyPar~\cite{kahypar0} adopted multilevel frameworks comprising coarsening, initial partitioning, and refinement phases, substantially improving both efficiency and quality~\cite{karypis1999multilevel}.

\begin{figure}[t]
    \centering
    \includegraphics[width=1\linewidth]{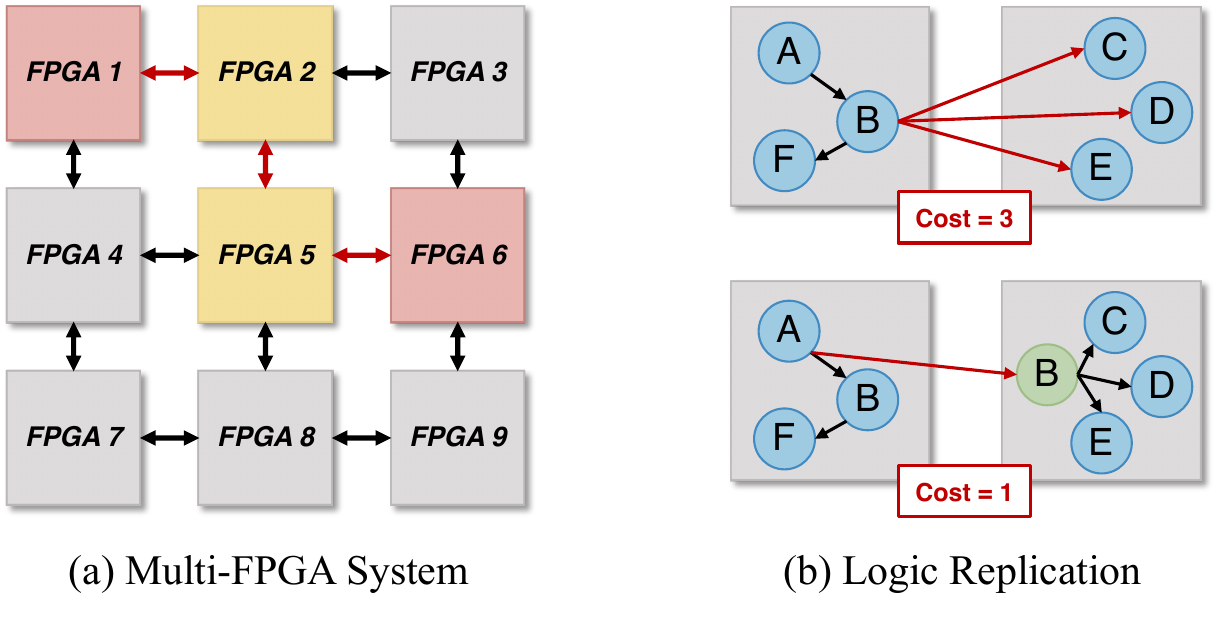}
    \vspace{-0.5cm}
    \caption{(a) Interconnect topology of a multi-FPGA system with heterogeneous resource constraints. (b) Communication reduction through strategic logic replication.}
    \label{background}
    \vspace{-0.3cm}
\end{figure}

However, hypergraph partitioning for MFS presents distinct characteristics compared to traditional approaches~\cite{748202hMetis, kahypar0}:

{\large \ding{182}} \textbf{Topology Awareness.} 
Traditional partitioning algorithms focus on minimizing cut-size, which is insufficient for MFS~\cite{Challenges10.1145/3177540.3177542}. 
Due to limited I/O resources, FPGAs form a specific topological network rather than a full-mesh, as shown in Fig.~\ref{background}(a). 
For non-adjacent FPGA pairs, signals must traverse intermediate FPGAs as hops, increasing transmission delays. 
Additionally, Time-Division-Multiplexing (TDM)~\cite{TDM640619, zou2020time} is adopted to multiplex signal paths over single wires. 
As communication increases, both TDM ratio and latency rise. 
Thus, this work focuses on minimizing total hop distance~\cite{MaPart2024, Challenges10.1145/3177540.3177542} to optimize performance.

{\large \ding{183}} \textbf{MFS-specific Constraints.}
FPGAs contain various hardware resources~\cite{kuon2008fpga}, including Flip-Flops (FFs), Look-Up Tables (LUTs), and Digital Signal Processors (DSPs), each with distinct capacity limits. 
Circuit nodes must be assigned without exceeding any resource limit on each FPGA. 
Furthermore, inter-FPGA communication relies on I/O pins~\cite{kuon2008fpga}, imposing additional constraints. 
Many applications also constrain the maximum hop count to meet timing requirements~\cite{MaPart2024}.

Recent works including TopoPart~\cite{TopoPart9643481}, MaPart~\cite{MaPart2024}, and EasyPart~\cite{Tong2024EasyPartAE} have extended traditional approaches by incorporating MFS constraints. 
These methods typically model FPGA resource constraints through an imbalance factor \(\epsilon\), leaving FPGA resources underutilized after partitioning. 
We address this limitation by introducing logic replication~\cite{1218957logicreplication} to maximize hardware utilization. 
As shown in Fig.~\ref{background}(b), replication reduces inter-FPGA communication by enabling local access to replicated nodes. 
However, integrating logic replication into multilevel partitioning requires careful balancing, as replicated nodes consume resources that may compromise subsequent refinement operations.

This paper presents \method, a fully customized multilevel hypergraph partitioning framework for MFS that integrates logic replication with topology-aware optimization. Our contributions are summarized as follows:

\begin{enumerate}
    \item We adapt logic replication for multilevel hypergraph partitioning in MFS environments with a fully customized implementation that natively supports MFS constraints.
    
    \item We propose three key innovations in the multilevel partitioning framework: (1) dynamic coarsening with FPGA-aware scoring, (2) heat-value guided assignment, and (3) logic replication and deletion supported refinement, collectively improving partitioning quality.
    
    \item We conducted extensive experiments to validate our approach, which demonstrates 52.3\% average reduction in total hop distance compared to state-of-the-art partitioners with 11.1\(\times\) speedup. Our solution also outperforms winners of the EDA Elite Challenge Contest~\cite{EDAContest}.
\end{enumerate}
\section{Preliminary}

\subsection{Multi-FPGA System and Modeling}

Our work targets hypergraph partitioning with logic replication under MFS constraints. The problem input comprises the FPGA network and the Design Under Test (DUT).

The FPGA network specifies the number of FPGAs and their bidirectional connectivity.
We define hop distance as the shortest path length between two FPGAs in the network topology. 
The DUT consists of circuit components and interconnecting nets, where signals propagate from source nodes to all drain nodes. We model the DUT connectivity as a hypergraph, where the partitioning solution directly determines the MFS deployment strategy. Table~\ref{Notation} summarizes key notations used throughout this work.

Our optimization objective minimizes inter-FPGA communication cost, quantified through total hop distance~\cite{EDAContest}:

\[
\text{hop\_distance}(e) = \sum_{\hat{v} \in N(e)} \text{hop}(\text{part}(\text{source}(e)), \hat{v})
\]

\[
\text{total\_hop\_distance} = \sum_{e \in E} w_e \cdot \text{hop\_distance}(e)
\]

This metric directly correlates with communication latency, bandwidth contention, and TDM ratio in MFS \cite{farooq2018inter, Challenges10.1145/3177540.3177542}. Logic replication, illustrated in Fig.~\ref{background}(b), provides a mechanism to reduce communication cost by replicating source nodes across multiple FPGAs. While increasing resource consumption, strategic replication eliminates inter-FPGA signal routing, as shown where replicating node B onto FPGA 2 removes three cut edges at the cost of one additional communication link to node A.

\subsection{Multilevel Hypergraph Partitioning Framework}

We adopt the multilevel partitioning paradigm used by state-of-the-art tools such as KaHyPar \cite{10.1145/3529090kahypar} and hMETIS \cite{748202hMetis}. This framework operates through three sequential phases:

\textbf{Coarsening:} The hypergraph undergoes iterative reduction through vertex aggregation, creating a hierarchy of progressively smaller hypergraphs. Strongly connected vertices merge into hypernodes, reducing problem complexity while preserving essential structural properties.

\textbf{Assignment:} A coarse-level hypergraph receives an initial partition using specialized algorithms. Unlike traditional partitioning into equivalent clusters, our MFS-specific task assigns hypernodes to heterogeneous FPGAs with distinct resource capacities and topological positions.

\textbf{Refinement:} The partition propagates back through the hierarchy to the original hypergraph. Local optimization techniques, including KL \cite{6771089KL} and FM \cite{1585498FM} algorithms, progressively improve solution quality while respecting resource and communication constraints at each uncoarsening level.

\begin{table}[!tb]
\caption{Terminology and Notation.}
\label{Notation}
\renewcommand{\arraystretch}{1.15}
\resizebox{\columnwidth}{!}{%
\begin{tabular}{|l|l|}
\hline
\textbf{Term} & \textbf{Definition} \\ \hline
\(H(V, E)\) & Hypergraph with vertex set \(V\) and hyperedge set \(E\) \\ \hline
\(v\), \(e\) & Circuit unit (vertex) and circuit net (hyperedge) \\ \hline
\(|e|\) & Number of vertices in hyperedge \(e\) \\ \hline
\(\omega_i(v)\) & Type-\(i\) resource usage of vertex \(v\) \\ \hline
\(\overline{C}_i\) & Average FPGA capacity for resource type \(i\) \\ \hline
\(w_e\) & Weight of net \(e\) (number of signals in the net) \\ \hline
\(\text{source}(e)\), \(\text{drain}(e)\) & Source vertices and drain vertices set of \(e\) \\ \hline
\(I(v)\) & Set of hyperedges containing vertex \(v\) \\ \hline
\(N(e)\) & Set of FPGAs containing drain vertices of \(e\) \\ \hline
\(\text{part}(v)\) & FPGA assignment of vertex \(v\) \\ \hline
\(G(\hat{V}, \hat{E})\) & MFS network with FPGA set \(\hat{V}\) and connection set \(\hat{E}\) \\ \hline
\(\hat{v}\) & An individual FPGA in \(\hat{V}\) \\ \hline
\(\text{hop}(\hat{u}, \hat{v})\) & Hop distance between FPGAs \(\hat{u}\) and \(\hat{v}\) \\ \hline
\(\alpha\), \(\epsilon\), \(\ell\) & Penalty exponent, imbalance factor, coarsening level \\ \hline
\end{tabular}%
}
\vspace{-0.2cm}
\end{table}
\section{Methodology}

\subsection{Overview}

\method~adopts a multilevel hypergraph partitioning framework comprising three phases: coarsening, assignment, and refinement. We develop a fully customized implementation tailored for multi-FPGA systems, addressing the limitations of conventional partitioners through three key innovations:

\textbf{Dynamic Coarsening with FPGA-Aware Scoring.} Traditional coarsening ignores multi-dimensional FPGA constraints, often producing infeasible solutions. We introduce a dynamic merging strategy that balances connectivity preservation with resource utilization, adapting penalty weights across coarsening levels to enable effective initial partition generation.

\textbf{Heat-Value Guided Assignment.} Unlike homogeneous partitioning, MFS assignment must consider heterogeneous resource constraints and network topology. We propose a heat-value metric that quantifies architectural significance of both FPGAs and nodes, coupled with a deep backtracking mechanism to explore diverse solution subspaces.

\textbf{Replication and Deletion Supported Refinement.} Beyond traditional move and exchange operations, we introduce replication to exploit spare FPGA resources and deletion to reclaim space consumed by coarse-level replications. These operations are coordinated through a \(1+3K\)-Heap structure with incremental gain updates for efficient optimization.

\subsection{Dynamic Coarsening with FPGA-Aware Scoring}

\begin{figure}
    \centering
    \includegraphics[width=1\linewidth]{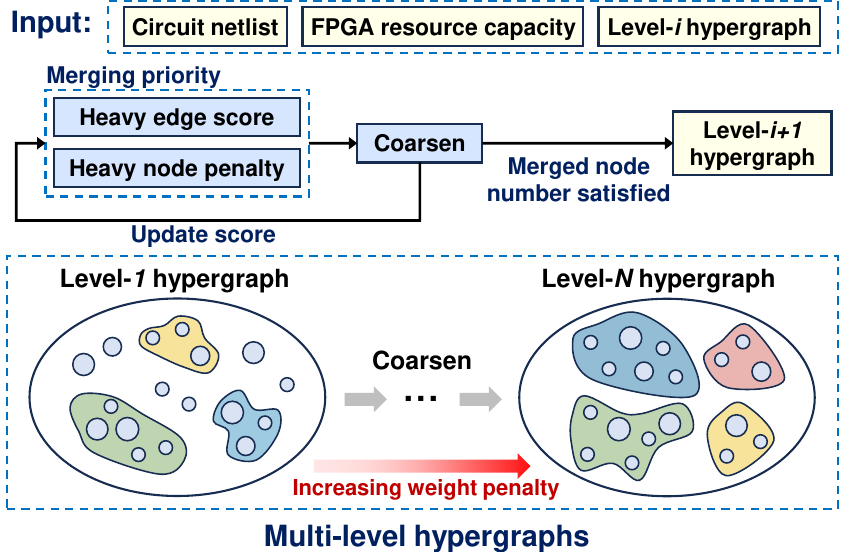}
    \caption{Dynamic Coarsening Process. 
    }
    \label{fig:dynamic coarsening}
    \vspace{-0.3cm}
\end{figure}

Coarsening merges strongly-connected nodes into hypernodes across multiple hierarchy levels, enabling later refinement at varying granularities. Our approach balances two objectives: preserving connectivity within clusters while maintaining balanced resource utilization for effective assignment.

\subsubsection{Connectivity and Balance Scoring}

We evaluate merging candidates using connectivity strength and resource balance. For node pair \((u, v)\), the \textbf{HeavyEdgeScore}~\cite{C2011patoh, 748202hMetis, 10.1145/3529090kahypar} measures connectivity:
\[
r(u, v) = \sum_{\substack{e \in I(v) \cap I(u)}} \frac{w_e}{|e| - 1}
\]
where \(I(v)\) denotes hyperedges containing \(v\), \(w_e\) is edge weight, and \(|e|\) is edge size. This metric prioritizes high-weight, low-degree connections.

The \textbf{HeavyNodePenalty} enforces multi-dimensional resource balance:
\[
p(u, v) = \left( \sum_{i=1}^k \frac{\omega_i(u) \cdot \omega_i(v)}{\overline{C}_i^2}  \right)^{\alpha}
\]
where \(\omega_i(v)\) represents type-\(i\) resource usage of node \(v\), \(\overline{C}_i\) is the average FPGA capacity for resource type \(i\), \(k\) is the number of resource types, and \(\alpha\) is the penalty exponent.

\subsubsection{Adaptive Penalty Adjustment}

Early coarsening should emphasize connectivity, while later stages require balanced clusters. We dynamically adjust \(\alpha\) across levels \(\ell\):
\[
\alpha = \alpha_0 + \frac{\Delta_\alpha \cdot \ln 2}{\ln\left( \frac{N_{\text{init}} + 1}{N_{\text{final}}} \right)} \cdot \ell
\]
where \(\alpha_0\) and \(\Delta_\alpha\) define baseline and increment parameters, \(N_{\text{init}}\) and \(N_{\text{final}}\) specify initial and final node counts. This logarithmic scaling smoothly transitions from connectivity-focused to balance-focused merging.

The final merging priority combines both criteria:
\[
\Psi(u,v) = \frac{r(u,v)}{p(u,v)}
\]

As illustrated in Fig.~\ref{fig:dynamic coarsening}, early levels create tightly-connected but uneven clusters (left), while later levels enforce balanced resource distribution (right). Our experiments show this adaptive framework improves initial solution quality by 18.7\% over static penalty methods, as shown in Sec.~\ref{subsec:ablation}.

\subsection{Heat-Value Guided Assignment}

\begin{figure}
    \centering
    \includegraphics[width=1\linewidth]{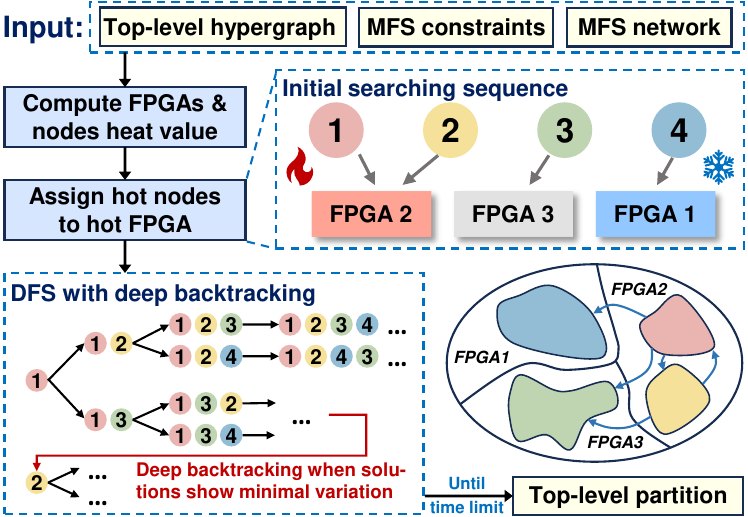}
    \caption{Heat-Value Guided Assignment Process. }
    \label{fig:assignment}
    \vspace{-0.3cm}
\end{figure}

Traditional partitioning treats all targets as homogeneous clusters, ignoring architectural heterogeneity. In MFS, each FPGA has distinct resource constraints and topological positions. We reformulate this as an architecture-aware assignment problem using heat-value metrics and deep backtracking.

\subsubsection{Heat-Value Metrics}

We quantify architectural significance of both FPGAs and nodes:

\textbf{FPGA Heat Value} evaluates topological centrality:
\[
\mathrm{Hop\_sum}(\hat{v}) = \sum_{\hat{u} \in \hat{V} \setminus \{\hat{v}\}} \begin{cases} 
\mathrm{hop}(\hat{v}, \hat{u}) & \text{if } \mathrm{hop} \leq \mathrm{Hop}_{\mathrm{max}} \\
\beta \cdot \mathrm{Hop}_{\mathrm{max}} & \text{otherwise}
\end{cases}
\]
\[
\mathrm{Heat}_{\mathrm{FPGA}}(\hat{v}) = \frac{\sum_{i=1}^{k} \omega_i(\hat{v})^2}{\mathrm{Hop\_sum}(\hat{v})}
\]
where \(\hat{v}\) denotes an FPGA, \(\hat{V}\) is the FPGA set, \(\mathrm{hop}(\hat{v}, \hat{u})\) is hop distance between FPGAs, \(\mathrm{Hop}_{\mathrm{max}}\) is the maximum hop constraint, and \(\beta\) is the penalty coefficient (set to 2).

\textbf{Node Heat Value} measures connectivity and resource criticality:
\[
\mathrm{Heat}_{\mathrm{node}}(v) = \sum_{e \in I(v)} w_e \cdot \sum_{i=1}^{k} \omega_i(v)^2
\]

High heat-value nodes are prioritized for placement on central FPGAs to minimize communication cost (Fig.~\ref{fig:assignment}).

\subsubsection{Deep Backtracking Exploration}

Assignment performs depth-first search with three pruning conditions: (1) exceeding current best total hop distance, (2) violating communication utilization, and (3) exceeding FPGA resource capacity.

Solution spaces contain multiple local minima regions. Conventional methods often start from an arbitrary initial partitioning, which confines the subsequent refinement to a single local optimum. Our key insight is that assignment should explore diverse solution subspaces rather than merely finding low-cost solutions. Solutions with better structural properties enable superior refinement outcomes.

As shown in Fig.~\ref{fig:assignment}, we implement deep backtracking that triggers when the THD of consecutive solutions show minimal variation (\(<2\%\) normalized cost difference). This mechanism performs aggressive pruning to escape local valleys, enabling cross-valley exploration. Heat-value scoring provides theoretical grounding: early assignment variations of high heat-value nodes fundamentally reshape solution trajectories, creating distinct subspaces with unique refinement potential.

\begin{figure}
    \centering
    \includegraphics[width=1\linewidth]{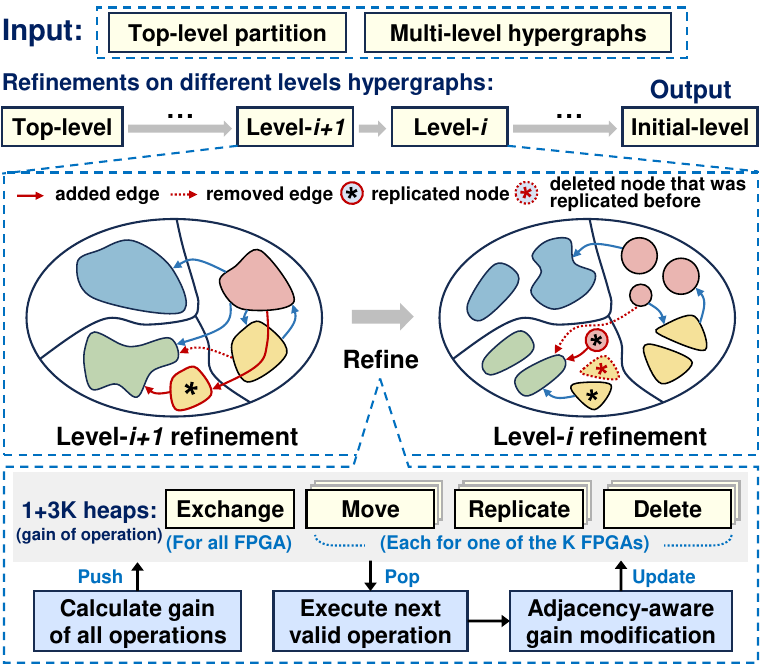}
    \caption{Logic Replication Supported Refinement Process.}
    \label{fig:refinement}
    \vspace{-0.3cm}
\end{figure}

\subsection{Logic Replication and Deletion Supported Refinement}

Refinement optimizes the initial partition through multi-level adaptation, leveraging hierarchical structural properties to progressively improve solution quality.

\subsubsection{Four-Operation Framework}

Beyond traditional \textbf{move} and \textbf{exchange} operations~\cite{10.1145/3529090kahypar}, we introduce two MFS-specific operations. The \textbf{replicate} operation replicates nodes across multiple FPGAs, utilizing remaining resources to reduce inter-FPGA communication by enabling local signal access. However, replicated nodes consume FPGA resources, constraining fine-grained refinement space.

The \textbf{delete} operation addresses this by removing unwanted replicated nodes to reclaim resources for other operations. Delete executes when its gain is zero or positive, prioritizing resource recovery over immediate gain to create optimization headroom for subsequent steps.

\subsubsection{Multi-Operator Coordination via \(1+3K\)-Heap}

We coordinate four operation types through a \(1+3K\)-Heap structure, as illustrated in Fig.~\ref{fig:refinement}. One heap tracks cross-FPGA node exchange opportunities. For move, replicate, and delete operations, we maintain \(K\) heaps for each operation type, where \(K\) is the number of FPGAs. Each heap stores operation gains when targeting a specific FPGA as the destination.

The algorithm iteratively applies the highest-gain operation from these heaps. Each operation execution triggers online updates to total hop distance, resource usage, and communication utilization. Operation gains are dynamically updated through adjacency-aware propagation, where node modifications trigger localized gain recalculations for neighboring hyperedges.

This incremental update mechanism, though algorithmically complex in multi-FPGA contexts, achieves significant speedup. Our implementation delivers an average \(2.27\times\) acceleration compared to naive full recalculation while maintaining identical solution quality, making refinement computationally efficient for large-scale problems.

\begin{table}[t]
  \caption{MFS configurations.}
  \label{tab:MFS_config}
\resizebox{\columnwidth}{!}{%
\begin{tabular}{|c|cccccc|}
\hline
\cellcolor[HTML]{EFEFEF}Testcase & case01 & case02 & case03 & case04 & case05      & case06     \\ \hline
\cellcolor[HTML]{EFEFEF}\#FPGAs  & 4      & 8      & 32     & 64     & 4           & 8          \\
\cellcolor[HTML]{EFEFEF}\#Links  & 3      & 11     & 88     & 173    & 3           & 28         \\
\cellcolor[HTML]{EFEFEF}\#Types  & 8      & 8      & 8      & 8      & 8           & 8          \\ \hline
\cellcolor[HTML]{EFEFEF}Testcase & case07 & case08 & case09 & case10 & SampleInput & synopsys02 \\ \hline
\cellcolor[HTML]{EFEFEF}\#FPGAs  & 32     & 32     & 64     & 64     & 8           & 56         \\
\cellcolor[HTML]{EFEFEF}\#Links  & 92     & 92     & 177    & 177    & 11          & 157        \\
\cellcolor[HTML]{EFEFEF}\#Types  & 8      & 8      & 8      & 8      & 1           & 1          \\ \hline
\end{tabular}
}
\end{table}

\begin{table}[t]
  \caption{Total hop distance comparison on the EDA Contest benchmark~\cite{EDAContest}.}
  \label{tab:eda_contest_results}
\renewcommand{\arraystretch}{1.2}
\resizebox{\columnwidth}{!}{%
\begin{tabular}{ccccccc}
\hline
\rowcolor[HTML]{EFEFEF}
Testcase & \#Nodes & \#Edges & KaHyPar & 2nd place & 1st place & \method \\ \hline
case01 & 16 & 13 & 17 & 11 & 8 & \textbf{8} \\
case02 & 600 & 1239 & 3608 & 2321 & 2456 & \textbf{2394} \\
case03 & 11451 & 31071 & 13909 & 9870 & 7371 & \textbf{6199} \\
case04 & 1000000 & 953817 & 57850 & 46194 & 19469 & \textbf{14077} \\
case05 & 1600 & 2157 & 299 & 182 & 147 & \textbf{118} \\
case06 & 1053 & 2447 & 409 & 347 & 300 & \textbf{300} \\
case07 & 40000 & 45504 & 7843 & 7654 & \textbf{5547} & 5702 \\
case08 & 25000 & 27482 & 6370 & 5898 & 4324 & \textbf{3631} \\
case09 & 240000 & 253232 & 46335 & 45720 & 33276 & \textbf{31090} \\
case10 & 900000 & 1108491 & 24241 & 27983 & 14859 & \textbf{12027} \\ \hline
\rowcolor[HTML]{C0C0C0}
Avg.ratio & -- & -- & 1.00 & 0.91 & 0.545 & 0.47 \\ \hline
\end{tabular}%
}
\vspace{-0.2cm}
\end{table}

\section{Experimental Results}

We implemented \method~in C++ and compiled it with g++ 11.4.0 on a machine equipped with an Intel Xeon PLATINUM 8558P 2.7-GHz CPU running Ubuntu 22.04.

\subsection{Experimental Setup}

\textbf{Hypergraphs.}
Experiments are conducted on two benchmarks. The first is Titan23~\cite{Titan10.1145/2629579}, a widely used suite of circuit netlists in which all hyperedges are unweighted and treated as equivalent. The second comprises 10 test cases from Problem 3 of the Integrated Circuit EDA Elite Challenge Contest~\cite{EDAContest}, where each hyperedge carries a weight. The node and edge counts for each benchmark are listed in the result tables.

\textbf{MFS Instances.}
For Titan23, we use two MFS instances from the ICCAD'19 Contest~\cite{ICCAD2019Contest8942051}: \textit{SampleInput} and \textit{synopsys02}. For the EDA Contest test cases, each case comes with its own unique FPGA system. These systems are substantially more complex: FPGAs are heterogeneous, each subject to eight distinct resource constraints (FF, LUT, BUFG, TBUF, DCM, BRAM, DSP, and PP), and each FPGA has an individual limit on the number of inter-FPGA connections. The MFS configurations are summarized in Table~\ref{tab:MFS_config}, where \#FPGAs denotes the number of FPGAs, \#Links denotes the number of inter-FPGA connections, and \#Types denotes the number of distinct resource constraint categories per FPGA.

\textbf{Constraints.}
All 10 EDA Contest cases impose strict multi-resource constraints on every FPGA.
For Titan23, the imbalance factor \(\epsilon\) is set to 0.2.

\subsection{Results Analysis}

\subsubsection{RePart-mini and RePart}

\submethod~follows the same multilevel pipeline as KaHyPar but is more concise and intuitive. It omits the topology-aware coarsening and heat-value guided assignment strategies proposed in this work, and only retains node replication in the refinement phase to isolate the effect of logic replication. \method~incorporates all proposed optimizations and achieves significant improvements in both total hop distance and runtime over KaHyPar.

\subsubsection{Total Hop Distance Reduction}

Table~\ref{tab:eda_contest_results} reports results on the EDA Contest benchmark, where total hop distance is the primary metric. \submethod, equipped solely with node replication, achieves an average total hop distance of 12540.4, which is 77.9\% of KaHyPar's 16088.1, confirming that replication alone yields a meaningful gain. \method~further reduces the average to 7554.6, only 47.0\% of KaHyPar's result. Compared to the contest winners, \method~achieves a 14.0\% reduction over the 1st-place solution and a 48.3\% reduction over the 2nd-place solution.

Table~\ref{tab:titan23_synopsys02} presents results on the Titan23 benchmark under the \textit{synopsys02} MFS with \(\epsilon = 0.2\). Although Titan23 is closer to a general partitioning problem due to its unweighted edges and single resource type, our methods still deliver strong results. \submethod~achieves a notable reduction in total hop distance over KaHyPar with significantly lower runtime. \method~reduces total hop distance by 52.3\% and runtime by 98.1\% compared to KaHyPar, demonstrating that the proposed optimizations are effective across both simple and complex MFS settings.

\begin{table}[t]
  \caption{Comparison on the Titan23 benchmark under the \textit{synopsys02} MFS. THD denotes total hop distance; \(t\) denotes runtime in seconds.}
  \centering
  \label{tab:titan23_synopsys02}
\renewcommand{\arraystretch}{1.3}
\resizebox{\columnwidth}{!}{
\begin{tabular}{|c|c|c|cc|cc|cc|}
\hline
\rowcolor[HTML]{EFEFEF}
& & &
\multicolumn{2}{c|}{\cellcolor[HTML]{EFEFEF}KaHyPar} &
\multicolumn{2}{c|}{\cellcolor[HTML]{EFEFEF}\submethod} &
\multicolumn{2}{c|}{\cellcolor[HTML]{EFEFEF}\method} \\ \cline{4-9}
\multirow{-2}{*}{\cellcolor[HTML]{EFEFEF}Testcase} &
\multirow{-2}{*}{\cellcolor[HTML]{EFEFEF}\#Nodes} &
\multirow{-2}{*}{\cellcolor[HTML]{EFEFEF}\#Edges} &
\multicolumn{1}{c|}{\cellcolor[HTML]{EFEFEF}THD} &
\multicolumn{1}{c|}{\cellcolor[HTML]{EFEFEF}t} &
\multicolumn{1}{c|}{\cellcolor[HTML]{EFEFEF}THD} &
\multicolumn{1}{c|}{\cellcolor[HTML]{EFEFEF}t} &
\multicolumn{1}{c|}{\cellcolor[HTML]{EFEFEF}THD} &
\multicolumn{1}{c|}{\cellcolor[HTML]{EFEFEF}t} \\ \hline

sparcT1\_core & 91976 & 92827 &
\multicolumn{1}{c|}{57806} & \multicolumn{1}{c|}{3419} &
\multicolumn{1}{c|}{32610} & \multicolumn{1}{c|}{\textbf{66}} &
\multicolumn{1}{c|}{\textbf{22391}} & \multicolumn{1}{c|}{81} \\ \hline

neuron & 92290 & 125305 &
\multicolumn{1}{c|}{10838} & \multicolumn{1}{c|}{2933} &
\multicolumn{1}{c|}{8546} & \multicolumn{1}{c|}{\textbf{48}} &
\multicolumn{1}{c|}{\textbf{6239}} & \multicolumn{1}{c|}{85} \\ \hline

stereo\_vision & 94050 & 12708 &
\multicolumn{1}{c|}{10514} & \multicolumn{1}{c|}{\textbf{25}} &
\multicolumn{1}{c|}{8835} & \multicolumn{1}{c|}{50} &
\multicolumn{1}{c|}{\textbf{6655}} & \multicolumn{1}{c|}{89} \\ \hline

des90 & 111221 & 139557 &
\multicolumn{1}{c|}{33170} & \multicolumn{1}{c|}{3113} &
\multicolumn{1}{c|}{18175} & \multicolumn{1}{c|}{\textbf{56}} &
\multicolumn{1}{c|}{\textbf{15457}} & \multicolumn{1}{c|}{66} \\ \hline

SLAM\_spheric & 113115 & 142408 &
\multicolumn{1}{c|}{54909} & \multicolumn{1}{c|}{633} &
\multicolumn{1}{c|}{32322} & \multicolumn{1}{c|}{\textbf{53}} &
\multicolumn{1}{c|}{\textbf{24967}} & \multicolumn{1}{c|}{157} \\ \hline

cholesky\_mc & 113250 & 144948 &
\multicolumn{1}{c|}{21987} & \multicolumn{1}{c|}{39} &
\multicolumn{1}{c|}{17818} & \multicolumn{1}{c|}{\textbf{27}} &
\multicolumn{1}{c|}{\textbf{16356}} & \multicolumn{1}{c|}{67} \\ \hline

segmentation & 138295 & 179051 &
\multicolumn{1}{c|}{28937} & \multicolumn{1}{c|}{241} &
\multicolumn{1}{c|}{14546} & \multicolumn{1}{c|}{\textbf{59}} &
\multicolumn{1}{c|}{\textbf{11563}} & \multicolumn{1}{c|}{84} \\ \hline

bitonic\_mesh & 192064 & 235328 &
\multicolumn{1}{c|}{29297} & \multicolumn{1}{c|}{2394} &
\multicolumn{1}{c|}{18746} & \multicolumn{1}{c|}{\textbf{77}} &
\multicolumn{1}{c|}{\textbf{12636}} & \multicolumn{1}{c|}{95} \\ \hline

dart & 202354 & 223301 &
\multicolumn{1}{c|}{24432} & \multicolumn{1}{c|}{2813} &
\multicolumn{1}{c|}{19237} & \multicolumn{1}{c|}{\textbf{61}} &
\multicolumn{1}{c|}{\textbf{14692}} & \multicolumn{1}{c|}{202} \\ \hline

openCV & 217453 & 284108 &
\multicolumn{1}{c|}{49959} & \multicolumn{1}{c|}{2582} &
\multicolumn{1}{c|}{20874} & \multicolumn{1}{c|}{\textbf{88}} &
\multicolumn{1}{c|}{\textbf{13376}} & \multicolumn{1}{c|}{220} \\ \hline

stap\_qrd & 240240 & 290123 &
\multicolumn{1}{c|}{31803} & \multicolumn{1}{c|}{312} &
\multicolumn{1}{c|}{23558} & \multicolumn{1}{c|}{\textbf{95}} &
\multicolumn{1}{c|}{\textbf{21289}} & \multicolumn{1}{c|}{143} \\ \hline

minres & 261359 & 320540 &
\multicolumn{1}{c|}{18326} & \multicolumn{1}{c|}{318} &
\multicolumn{1}{c|}{16427} & \multicolumn{1}{c|}{\textbf{84}} &
\multicolumn{1}{c|}{\textbf{11347}} & \multicolumn{1}{c|}{148} \\ \hline

cholesky\_bdti & 266422 & 342688 &
\multicolumn{1}{c|}{38213} & \multicolumn{1}{c|}{278} &
\multicolumn{1}{c|}{27886} & \multicolumn{1}{c|}{\textbf{98}} &
\multicolumn{1}{c|}{\textbf{25046}} & \multicolumn{1}{c|}{163} \\ \hline

denoise & 275638 & 356848 &
\multicolumn{1}{c|}{18765} & \multicolumn{1}{c|}{801} &
\multicolumn{1}{c|}{15269} & \multicolumn{1}{c|}{\textbf{69}} &
\multicolumn{1}{c|}{\textbf{11049}} & \multicolumn{1}{c|}{79} \\ \hline

sparcT2\_core & 300109 & 302663 &
\multicolumn{1}{c|}{105303} & \multicolumn{1}{c|}{608} &
\multicolumn{1}{c|}{46209} & \multicolumn{1}{c|}{\textbf{156}} &
\multicolumn{1}{c|}{\textbf{43738}} & \multicolumn{1}{c|}{405} \\ \hline

gsm\_switch & 493260 & 507821 &
\multicolumn{1}{c|}{151638} & \multicolumn{1}{c|}{7721} &
\multicolumn{1}{c|}{69345} & \multicolumn{1}{c|}{\textbf{259}} &
\multicolumn{1}{c|}{\textbf{46650}} & \multicolumn{1}{c|}{469} \\ \hline

mes\_noc & 547544 & 577664 &
\multicolumn{1}{c|}{33654} & \multicolumn{1}{c|}{2606} &
\multicolumn{1}{c|}{21982} & \multicolumn{1}{c|}{\textbf{224}} &
\multicolumn{1}{c|}{\textbf{16817}} & \multicolumn{1}{c|}{379} \\ \hline

LU\_Network & 635456 & 726999 &
\multicolumn{1}{c|}{\textbf{44283}} & \multicolumn{1}{c|}{8206} &
\multicolumn{1}{c|}{50429} & \multicolumn{1}{c|}{\textbf{284}} &
\multicolumn{1}{c|}{52066} & \multicolumn{1}{c|}{607} \\ \hline

LU230 & 574372 & 669477 &
\multicolumn{1}{c|}{81471} & \multicolumn{1}{c|}{13659} &
\multicolumn{1}{c|}{53387} & \multicolumn{1}{c|}{\textbf{260}} &
\multicolumn{1}{c|}{\textbf{41990}} & \multicolumn{1}{c|}{727} \\ \hline

sparcT1\_chip2 & 820886 & 821274 &
\multicolumn{1}{c|}{98869} & \multicolumn{1}{c|}{14491} &
\multicolumn{1}{c|}{45129} & \multicolumn{1}{c|}{\textbf{442}} &
\multicolumn{1}{c|}{\textbf{36679}} & \multicolumn{1}{c|}{853} \\ \hline

directrf & 931275 & 1374742 &
\multicolumn{1}{c|}{67096} & \multicolumn{1}{c|}{1770} &
\multicolumn{1}{c|}{33196} & \multicolumn{1}{c|}{\textbf{355}} &
\multicolumn{1}{c|}{\textbf{24895}} & \multicolumn{1}{c|}{509} \\ \hline

bitcoin\_miner & 1089284 & 1448151 &
\multicolumn{1}{c|}{89160} & \multicolumn{1}{c|}{5149} &
\multicolumn{1}{c|}{66531} & \multicolumn{1}{c|}{\textbf{451}} &
\multicolumn{1}{c|}{\textbf{48561}} & \multicolumn{1}{c|}{1060} \\ \hline

\rowcolor[HTML]{C0C0C0}
Avg.ratio & \cellcolor[HTML]{C0C0C0} & \cellcolor[HTML]{C0C0C0} &
\multicolumn{1}{c|}{\cellcolor[HTML]{C0C0C0}1} &
\multicolumn{1}{c|}{\cellcolor[HTML]{C0C0C0}1} &
\multicolumn{1}{c|}{\cellcolor[HTML]{C0C0C0}0.600} &
\multicolumn{1}{c|}{\cellcolor[HTML]{C0C0C0}0.045} &
\multicolumn{1}{c|}{\cellcolor[HTML]{C0C0C0}0.477} &
\multicolumn{1}{c|}{\cellcolor[HTML]{C0C0C0}0.090} \\ \hline

\end{tabular}%
}
\vspace{-0.3cm}
\end{table}

\subsubsection{Cut Size Comparison}
Prior works TopoPart~\cite{TopoPart9643481}, Li et al.~\cite{AnEfficient10546878}, and RepPart~\cite{Wu2025RepPartAE} use cut size as their primary evaluation metric. However, cut size does not reflect the actual topological communication cost in an MFS when the maximum hop count exceeds one, since it treats all inter-FPGA edges equally regardless of the physical distance between FPGAs. Total hop distance captures this topology-dependent cost and is therefore a more faithful metric for real MFS deployments~\cite{EDAContest}. Because these prior methods do not support weighted hyperedges and are not open-source, we compare against them only on the Titan23 \textit{SampleInput} instance using each paper's originally reported figures.

As shown in Table~\ref{tab:titan23_sampleinput}, \method~achieves a smaller cut size than all baselines even though cut size is not its optimization objective. This result indicates that optimizing for total hop distance simultaneously yields high-quality partitions from the perspective of traditional cut-based metrics.

\begin{table}[t]
  \caption{Cut size comparison on the Titan23 benchmark under the \textit{SampleInput} MFS.}
  \label{tab:titan23_sampleinput}
\renewcommand{\arraystretch}{1.2}
\resizebox{\columnwidth}{!}{%
\begin{tabular}{ccccccc}
\hline
\rowcolor[HTML]{EFEFEF}
Testcase & \#Nodes & \#Nets & TopoPart & Li.~\cite{AnEfficient10546878} & RepPart & RePart \\ \hline
sparcT1\_core   & 91976   & 92827   & 54841  & 1048  & 1185   & 4255  \\
neuron          & 92290   & 125305  & 8815   & 4384   & 1197   & \textbf{711}  \\
stereo\_vision  & 94050   & 12708   & 32068  & 388   & 1110   & 723   \\
des90           & 111221  & 139557  & 87211  & 656   & 745    & 1221  \\
SLAM\_spheric   & 113115  & 142408  & 54402  & 7642   & 6987   & \textbf{6691} \\
cholesky\_mc    & 113250  & 144948  & 37658  & 428   & 1004   & 1686  \\
segmentation    & 138295  & 179051  & 34092  & 982   & 1509   & \textbf{940}   \\
bitonic\_mesh   & 192064  & 235328  & 184894 & 1169   & 1078 & 1893  \\
dart            & 202354  & 223301  & 168197 & 2776   & 2753   & \textbf{2167} \\
openCV          & 217453  & 284108  & 120428 & 525   & 1047   & 766   \\
stap\_qrd       & 240240  & 290123  & 134256 & 6049   & 3193   & \textbf{1648} \\
minres          & 261359  & 320540  & 96178  & 13617  & 4512   & \textbf{732}  \\
cholesky\_bdti  & 266422  & 342688  & 75479  & 578    & 572  & 1645  \\
denoise         & 275638  & 356848  & 93464  & 595   & 859    & 1023  \\
sparcT2\_core   & 300109  & 302663  & 98128  & 4920   & 1023 & 4500  \\
gsm\_switch     & 493260  & 507821  & 267424 & 14999  & 10191  & \textbf{7390} \\
mes\_noc        & 547544  & 577664  & 297140 & 324    & 311  & 3405  \\
LU230           & 574372  & 669477  & 134877 & 269   & 320    & 4878  \\
LU\_Network     & 635456  & 726999  & 152508 & 32467  & 9915   & \textbf{5100} \\
sparcT1\_chip2  & 820886  & 821274  & 207814 & 1417   & 958  & 2283  \\
directrf        & 931275  & 1374742 & 113493 & 435   & 936    & 2249  \\
bitcoin\_miner  & 1089284 & 1448151 & 413566 & 43935  & 9345   & \textbf{1125} \\ \hline
\rowcolor[HTML]{C0C0C0}
Avg.ratio & -- & -- & 1 & 0.049 & 0.022 & 0.019 \\
\hline
\end{tabular}%
}
\vspace{-0.3cm}
\end{table}

\subsection{Ablation Study}
\label{subsec:ablation}

We conduct an ablation study on the 10 EDA Contest test cases to evaluate the individual contribution of each proposed technique. Fig.~\ref{fig:ablation_study} reports the normalized total hop distance across three subfigures, each corresponding to one phase of the multilevel pipeline. The red bars represent the configurations that incorporate the proposed innovation for that phase.

\subsubsection{Coarsening}

\begin{figure}[t]
  \centering
  \includegraphics[width=1\columnwidth]{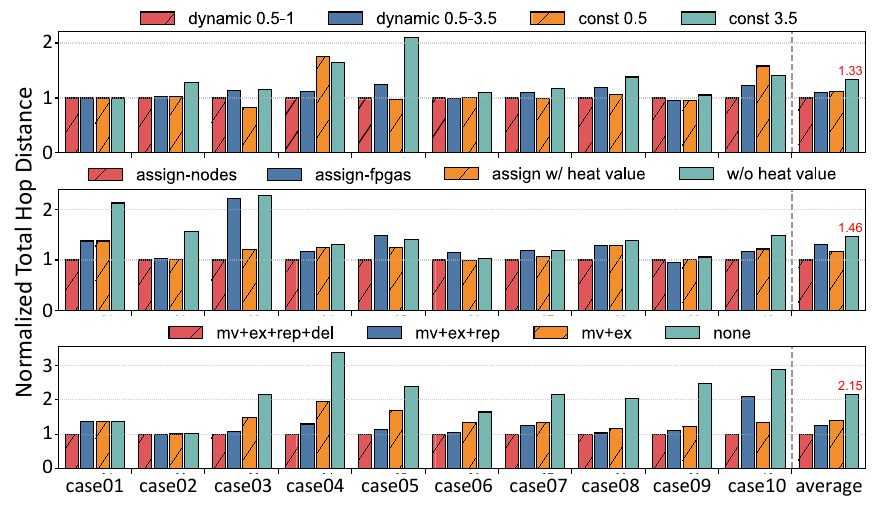}
  \caption{Ablation study on the EDA Contest benchmark. Normalized total hop distance is shown for coarsening (top), assignment (middle), and refinement (bottom).
  }
  \label{fig:ablation_study}
    \vspace{-0.3cm}
\end{figure}

The first subfigure in Fig.~\ref{fig:ablation_study} examines the effect of the adaptive penalty exponent \(\alpha\) in the FPGA-aware coarsening score. Using a fixed value of \(\alpha = 3.5\) yields an average total hop distance 20\% higher than the dynamic schedule that gradually increases \(\alpha\) from 0.5 to 3.5, confirming that the adaptive range is important for guiding early-stage coarsening decisions.

\subsubsection{Assignment}

The second subfigure compares three variants of the heat-value guided assignment. \textit{assign-nodes} uses four threads, each emphasizing differences among nodes during heat score evaluation. \textit{assign-fpgas} also uses four threads but focuses instead on differences among FPGAs. \textit{assign} is the single-threaded baseline. Since node-level variability is typically more pronounced in practice, \textit{assign-nodes} consistently achieves the best results and is adopted in \method.

\subsubsection{Refinement}

The third subfigure evaluates the contribution of each refinement operation: move (\textit{mv}), exchange (\textit{ex}), replication (\textit{rep}), and deletion (\textit{del}). Applying all four operations achieves a \(2.15\times\) reduction in average total hop distance compared to no refinement, and a \(1.39\times\) reduction compared to the \textit{mv+ex}-only baseline. Case 10 highlights the necessity of the deletion operation: \textit{mv+ex+rep} alone performs worse than \textit{mv+ex} in this case because unconstrained replication occupies FPGA resources and limits the subsequent move and exchange operations. The deletion operation releases these resources, enabling further optimization and recovering the performance loss.

\section{Conclusion}

This paper presents \method, an open-source hypergraph partitioning framework that integrates logic replication with topology-aware optimization across all phases of the partitioning pipeline. \method~effectively reduces inter-FPGA communication while exploiting available hardware resources. Experimental results demonstrate that \method~reduces total hop distance by 52.3\% on average compared to state-of-the-art hypergraph partitioning algorithms, while requiring only 9.0\% of its runtime. \method~also outperforms the winners of the EDA Elite Challenge Contest,
confirming its effectiveness on both standard and highly constrained MFS benchmarks.
\section*{ACKNOWLEDGMENTS}
We sincerely thank Jiarui Wang, Zizheng Guo, and Benzheng Li for their valuable comments and constructive guidance in improving this manuscript. We also gratefully acknowledge the Organizing Committee of the China Postgraduate IC Innovation Competition \textperiodcentered{} EDA Elite Challenge Contest and S2C for providing the datasets for this work.

\bibliographystyle{IEEEtran}
\bibliography{main}

\end{document}